\documentclass{article}

\usepackage{amsmath}
\usepackage{amssymb}
\usepackage{amsfonts}
\usepackage{bm}

\begin{document}


\newcommand{\tr}{\operatorname{Tr}}



\title{\bf Stress Tensor of Single Rigid Dumbbell by Virtual Work Method}
\author{Takashi Uneyama${}^{*,**,\dagger}$,
Tatsuma Oishi${}^{**}$,
Takato Ishida${}^{**}$, \\
Yuya Doi${}^{**}$, and
Yuichi Masubuchi${}^{**}$\\
\\
${}^{*}$JST-PRESTO and ${}^{**}$Department of Materials Physics, \\
Graduate School of Engineering, 
Nagoya University, \\
Furo-cho, Chikusa, Nagoya 464-8603, Japan}

\date{}

\maketitle

\begin{abstract}
We derive the stress tensor of a rigid dumbbell by using the virtual
work method. In the virtual work method, we virtually apply a small
deformation to the system, and relate the change of the energy to the
work done by the stress tensor. A rigid dumbbell consists of two
particles connected by a rigid bond of which length is constant
(the rigid constraint). The energy of the rigid dumbbell consists only on the kinetic energy.
Also, only the deformations which do not violate
the rigid constraint are allowed. Thus we need
the dynamic equations which is consistent with the rigid constraint
to apply the virtual deformation.
We rewrite the dynamic equations for the underdamped SLLOD-type dynamic equations
into the forms which are consistent with the rigid constraint.
Then we apply the virtual deformation
to a rigid dumbbell based on the obtained dynamic equations. We derive
the stress tensor for the rigid dumbbell model from the change of the kinetic energy.
Finally, we take the overdamped limit and
derive the stress tensor and the dynamic equation for the overdamped rigid dumbbell.
We show that the Green-Kubo type linear response formula can be reproduced by
combining the stress tensor and the dynamic equation at the overdamped limit.
\end{abstract}

%

\section{INTRODUCTION}

The stress tensor is one of the most important quantities in rheology.
A macroscopic material consists of molecules, and thus the stress tensor
can be related to the microscopic state of molecules. For example,
the stress tensor of a polymer melt can be related to the conformation
of polymer chains. By combining the molecular-level expression of the
stress tensor and the linear response theory, we can study the microscopic
molecular dynamics from the macroscopic rheological quantities.

There are several methods to calculate the stress tensor of a molecular
system. A simple yet useful method is so-called the method of the virtual
work\cite{Doi-Edwards-book,Doi-1987}. We virtually apply small deformation to the system, and relate the
change of the energy to the work done by the stress.
Here we briefly review the virtual work method.
For simplicity, we consider a dilute dumbbell suspension in which individual
dumbbell molecules do not interact each other.
In many cases, we can safely assume that the momentum relaxation is much
faster than the characteristic time scale of the bond vector (typically the orientation relaxation time), and thus we can
take the overdamped limit.
We model a dumbbell molecule
by two particles of which positions are expressed as $\bm{R}_{1}$ and $\bm{R}_{2}$.
In this work, we consider two types of dumbbell models: the rigid dumbbell and the flexible dumbbell.
In the rigid dumbbell model, two particles are connected by a rigid bond.
In the flexible dumbbell model,
two particles are connected by the tethering potential $\phi(\bm{r})$ with
$\bm{r} \equiv \bm{R}_{2} - \bm{R}_{1}$ being the bond vector.

We consider to virtually change the particle positions 
from $\bm{R}_{j}$ to $\bm{R}_{j}' = \bm{R}_{j} + \bm{E} \cdot \bm{R}_{j}$, where $\bm{E}$ 
is the displacement gradient tensor. Then the bond vector is changed to
$\bm{r}' = \bm{r} + \bm{E} \cdot \bm{r}$.
If the deformation is sufficiently small, the change of the potential energy
can be expressed as
\begin{equation}
 \label{tethering_energy_change_virtual_deformation}
 \phi(\bm{r}') - \phi(\bm{r})
  =  \frac{\partial \phi(\bm{r})}{\partial \bm{r}} \bm{r} : \bm{E} + O(\|\bm{E}\|^{2}),
\end{equation}
where $:$ represents the dyadic product (for two second order tensors $\bm{A}$ and $\bm{B}$, $\bm{A}:\bm{B} \equiv \sum_{\alpha\beta} A_{\alpha\beta} B_{\alpha\beta}$)
and $\| \bm{E} \|$ represents the norm of $\bm{E}$.
(Although there are several different definitions for the norm, we employ the $2$-norm in this work.)
The change of the energy can be interpreted as the work done by the 
stress, $V \hat{\bm{\sigma}}_{\text{bond}}^{\text{(OD)}}(\bm{r}) : \bm{E}$ with $V$ being the
volume of the system and $\hat{\bm{\sigma}}^{\text{(OD)}}_{\text{bond}}(\bm{r})$ being the stress tensor
by a single bond. (The superscript ``(OD)'' means the overdamped limit where
the momenta are assumed to be fully equilibrated.)
Then we have the single bond stress tensor as
\begin{equation}
 \label{stress_tensor_dumbbell_bond_kramers}
 \hat{\bm{\sigma}}_{\text{bond}}^{\text{(OD)}}(\bm{r}) = \frac{1}{V}  \frac{\partial \phi(\bm{r})}{\partial \bm{r}} \bm{r}.
\end{equation}

Although we can utilize the virtual work method in most cases,
it is not clear whether it is still applicable to systems with rigid constraints.
In the rigid dumbbell model, the bond length $|\bm{r}|$ is set to be
constant and there is no tethering potential. We cannot consider the change
of the potential energy by the virtual deformation. In addition, we cannot apply a deformation
which violates the rigid constraint. (One may consider that the rigid dumbbell
can be handled as the limit of the stiff tethering potential.
We will discuss the dumbbell model with the stiff tethering potential in Appendix~\ref{dumbbell_model_with_stiff_harmonic_dumbbell}.)

In this work, we show that we can derive the stress tensor for a rigid dumbbell
by the virtual work method. To apply the virtual work method to
the rigid dumbbell, we first derive the dynamic equations for the rigid dumbbell
under flow.
We employ the SLLOD-type dynamic equations for the bond vector and the bond momentum,
and rewrite the dynamic equations which are consistent with the rigid constraint.
Second, we apply the virtual deformation to the system by using the obtained
dynamic equations. This procedure gives the explicit expression for the
stress tensor of the rigid dumbbell in the underdamped system.
By taking the overdamped limit, we have the stress tensor for the rigid
dumbbell. Finally we calculate the shear relaxation modulus of a dilute rigid
dumbbell suspension.
The stress tensor at the overdamped limit
together with the dynamic equation at the overdamped limit can reproduce
the shear relaxation modulus calculated by the kinetic theory, except the instantaneous
delta function type term.

\section{MODEL}

\subsection{Flexible Dumbbell with Tethering Potential}

The energy of a single rigid dumbbell consists only on the kinetic energy.
Then, it would be reasonable for us to consider the contribution of the
kinetic energy to the stress tensor. Before we consider the rigid dumbbell,
here we consider the flexible dumbbell model with the tethering potential. We consider
the underdamped dynamics\cite{Uneyama-Nakai-Masubuchi-2019} and introduce
the momenta of two particles as $\bm{P}_{1}$ and $\bm{P}_{2}$.
The Hamiltonian of a single dumbbell is
\begin{equation}
 \label{hamiltonian_dumbbell}
 \mathcal{H}(\bm{R}_{1},\bm{R}_{2},\bm{P}_{1},\bm{P}_{2})
  = \frac{\bm{P}_{1}^{2} + \bm{P}_{2}^{2}}{2 M} + \phi(\bm{R}_{2} - \bm{R}_{1}),
\end{equation}
where $M$ is the mass of a particle.
To apply the
virtual deformation to the system, we consider the dynamic equations under an externally imposed flow.
We employ the SLLOD dynamic equations which
describe the dynamics of particles under a flow\cite{Evans-Morriss-book,Evans-Morriss-1984}.
By combining the SLLOD dynamic equations and the Langevin thermostat, the
dynamic equations become:
\begin{align}
 \label{sllod_dynamic_equation_tethering_potential_position}
 \frac{d\bm{R}_{j}(t)}{dt} & = \frac{1}{M} \bm{P}_{j}(t) + \bm{\kappa}(t) \cdot \bm{R}_{j}(t), \\
 \label{sllod_dynamic_equation_tethering_potential_momentum}
 \frac{d\bm{P}_{j}(t)}{dt} & = - \frac{\partial \phi(\bm{R}_{1}(t) - \bm{R}_{2}(t))}{\partial \bm{R}_{j}(t)}
 - \bm{\kappa}(t) \cdot \bm{P}_{j}(t)
 - \frac{Z}{M} \bm{P}_{j}(t) + \sqrt{2 Z k_{B} T} \bm{W}_{j}(t).
\end{align}
Here, $\bm{\kappa}(t) \equiv [\nabla \bm{v}(t)]^{\mathrm{T}}$ is the velocity gradient tensor
with $\bm{v}(t)$ being the imposed flow field, 
$Z$ is the friction coefficient of a particle, $k_{B}$ is the Boltzmann constant,
and $T$ is the temperature. $\bm{W}_{j}(t)$ is the Gaussian white
noise which satisfies the following relations:
\begin{equation}
 \label{fluctuation_dissipation_relation_tethering_potential}
 \langle \bm{W}_{j}(t) \rangle = 0, \qquad
  \langle \bm{W}_{j}(t) \bm{W}_{k}(t') \rangle = \delta_{jk} \bm{1} \delta(t - t'),
\end{equation}
where $\langle \dots \rangle$ represents the statistical average and $\bm{1}$ is the unit tensor.
The dynamic equations \eqref{sllod_dynamic_equation_tethering_potential_position} and \eqref{sllod_dynamic_equation_tethering_potential_momentum} can be decomposed into two set of statistically
independent equations.
We introduce the center of mass position $\bar{\bm{R}}(t) \equiv [\bm{R}_{1}(t) + \bm{R}_{2}(t)] / 2$
and the bond vector $\bm{r}(t) \equiv \bm{R}_{2}(t) - \bm{R}_{1}(t)$\cite{Uneyama-Nakai-Masubuchi-2019}.
We also introduce the momenta for the center of mass and the bond vector, $\bar{\bm{P}}(t) \equiv \bm{P}_{1}(t) + \bm{P}_{2}(t)$ and $\bm{p}(t) \equiv [\bm{P}_{2}(t) - \bm{P}_{1}(t)] / 2$.
Then the Hamiltonian \eqref{hamiltonian_dumbbell} can be rewritten as a function of $\bar{\bm{R}}$, $\bar{\bm{P}}$, $\bm{r}$, and $\bm{p}$:
\begin{equation}
 \label{hamiltonian_dumbbell_cm_bond}
 \mathcal{H}(\bar{\bm{R}},\bar{\bm{P}},\bm{r},\bm{p})
  = \frac{\bar{\bm{P}}^{2}}{2 \bar{M}}
  + \frac{\bm{p}^{2}}{2 m} + \phi(\bm{r}),
\end{equation}
where $\bar{M} \equiv 2 M$ and $m \equiv M / 2$ are masses.
Eqs~\eqref{sllod_dynamic_equation_tethering_potential_position} and
\eqref{sllod_dynamic_equation_tethering_potential_momentum} can be rewritten as
\begin{align}
 \label{sllod_dynamic_equation_tethering_potential_cm}
 \frac{d\bar{\bm{R}}(t)}{dt} & = \frac{1}{\bar{M}} \bar{\bm{P}}(t)
  + \bm{\kappa}(t) \cdot \bar{\bm{R}}(t), \\
 \label{sllod_dynamic_equation_tethering_potential_cm_momentum}
 \frac{d\bar{\bm{P}}(t)}{dt} & = 
 - \bm{\kappa}(t) \cdot \bar{\bm{P}}(t)
 - \frac{\bar{Z}}{\bar{M}} \bm{P}_{j}(t) + \sqrt{2 \bar{Z} k_{B} T} \bm{W}(t), \\
 \label{sllod_dynamic_equation_tethering_potential_bond}
 \frac{d\bm{r}(t)}{dt} & = \frac{1}{m} \bm{p}(t)
  + \bm{\kappa}(t) \cdot \bm{r}(t), \\
 \label{sllod_dynamic_equation_tethering_potential_bond_momentum}
 \frac{d\bm{p}(t)}{dt} & = - \frac{\partial \phi(\bm{r}(t))}{\partial \bm{r}(t)}
 - \bm{\kappa}(t) \cdot \bm{p}(t)
  - \frac{\zeta}{m} \bm{p}(t) + \sqrt{2 \zeta k_{B} T} \bm{w}(t).
\end{align}
Here, $\bar{Z} = 2 Z$ and $\zeta = Z / 2$
are the friction coefficients, and
$\bar{\bm{W}}(t) \equiv [\bm{W}_{1}(t) + \bm{W}_{2}(t)] / \sqrt{2}$ and
$\bm{w}(t) \equiv [\bm{W}_{2}(t) - \bm{W}_{1}(t)] / \sqrt{2}$ are Gaussian white
noises. It is straightforward to show that two Gaussian noises $\bar{\bm{W}}(t)$ and $\bm{w}(t)$ are statistically independent.
From eqs~\eqref{sllod_dynamic_equation_tethering_potential_cm}-\eqref{sllod_dynamic_equation_tethering_potential_bond_momentum},
we find that the dynamics of the center of mass $\bar{\bm{R}}(t)$ and the bond vector $\bm{r}(t)$
are statistically independent.

\subsection{Virtual Work Method}

We consider to apply the impulsive small strain to the system.
This can be done by setting the velocity gradient tensor 
in eqs~\eqref{sllod_dynamic_equation_tethering_potential_cm}-\eqref{sllod_dynamic_equation_tethering_potential_bond_momentum}
as an impulse at $t = 0$: $\bm{\kappa}(t) = \bm{E} \delta(t)$
As before, we assume that $\bm{E}$ is sufficiently small.
The positions and momenta change instantaneously around $t = 0$.
We express the center of mass positions and momenta just before and
just after the impulse as $\bar{\bm{R}} = \bar{\bm{R}}(-0)$,
$\bar{\bm{P}} = \bar{\bm{P}}(-0)$, $\bar{\bm{R}}' = \bar{\bm{R}}(+0)$,
and $\bar{\bm{P}}' = \bar{\bm{P}}(+0)$. In a similar way,
we express the bond vectors and the bond momenta before and after the impulse
as $\bm{r} = \bm{r}(-0)$, $\bm{p} = \bm{p}(-0)$,
$\bm{r}' = \bm{r}(+0)$, and $\bm{p}' = \bm{p}(+0)$. By integrating
the dynamic equations
\eqref{sllod_dynamic_equation_tethering_potential_cm}-\eqref{sllod_dynamic_equation_tethering_potential_bond_momentum}
from $t = -0$ to $t = +0$, we have
\begin{align}
 \label{virtual_deformation_cm}
\bar{\bm{R}}' & =  \bar{\bm{R}} + \bm{E} \cdot  \bar{\bm{R}}, &
  \bar{\bm{P}}' & =  \bar{\bm{P}} - \bm{E} \cdot  \bar{\bm{P}}, \\ 
 \label{virtual_deformation_bond}
 \bm{r}' & =  \bm{r} + \bm{E} \cdot \bm{r}, &
  \bm{p}' & =  \bm{p} - \bm{E} \cdot \bm{p}.
\end{align}
With the virtual deformation described by eqs~\eqref{virtual_deformation_cm}
and \eqref{virtual_deformation_bond}, the total energy of the dumbbell is changed
from $\mathcal{H}(\bar{\bm{R}},\bar{\bm{P}},\bm{r},\bm{p})$
to $\mathcal{H}(\bar{\bm{R}}',\bar{\bm{P}}',\bm{r}',\bm{p}')$.
The change of the energy can be related to the single dumbbell stress tensor as:
\begin{equation}
 \mathcal{H}(\bar{\bm{R}}',\bar{\bm{P}}',\bm{r}',\bm{p}')
  - \mathcal{H}(\bar{\bm{R}},\bar{\bm{P}},\bm{r},\bm{p})
  = V \hat{\bm{\sigma}}^{\text{(UD)}}(\bar{\bm{R}},\bar{\bm{P}},\bm{r},\bm{p}) : \bm{E}
  + O(\|\bm{E}\|^{2}),
\end{equation}
where $\hat{\bm{\sigma}}^{\text{(UD)}}(\bar{\bm{R}},\bar{\bm{P}},\bm{r},\bm{p})$
is the stress tensor of a single dumbbell in the underdamped system.
(The superscript ``(UD)'' means the underdamped system.)
The single dumbbell stress tensor can be decomposed into
the contributions of the center of mass and the bond vector as\cite{Uneyama-Nakai-Masubuchi-2019}:
\begin{align}
 \label{stress_tensor_dumbbell_dumbbell}
 \hat{\bm{\sigma}}^{\text{(UD)}}(\bar{\bm{R}},\bar{\bm{P}},\bm{r},\bm{p})
 & = \hat{\bm{\sigma}}_{\text{CM}}^{\text{(UD)}}(\bar{\bm{R}},\bar{\bm{P}})
 + \hat{\bm{\sigma}}_{\text{bond}}^{\text{(UD)}}(\bm{r},\bm{p}), \\
 \label{stress_tensor_dumbbell_cm}
 \hat{\bm{\sigma}}_{\text{CM}}^{\text{(UD)}}(\bar{\bm{R}},\bar{\bm{P}})
 & \equiv - \frac{1}{V} \frac{\bar{\bm{P}}\bar{\bm{P}}}{\bar{M}}, \\
 \label{stress_tensor_dumbbell_bond}
  \hat{\bm{\sigma}}_{\text{bond}}^{\text{(UD)}}(\bm{r},\bm{p})
 & \equiv \frac{1}{V} \left[ \frac{\partial \phi(\bm{r})}{\partial \bm{r}} \bm{r} - \frac{\bm{p}\bm{p}}{m} \right].
\end{align}

If the momentum relaxation process
is sufficiently fast, we can take the local equilibrium average over
$\bar{\bm{P}}$ and $\bm{p}$. This gives the stress tensor at the overdamped limit.
The local equilibrium distribution functions for momenta $\bar{\bm{P}}$ and $\bm{p}$ are
independent of $\bar{\bm{R}}$ and $\bm{r}$:
\begin{align}
 \Psi_{\text{eq}}(\bar{\bm{P}} | \bar{\bm{R}})
 & = 
  \left(\frac{1}{2 \pi \bar{M} k_{B} T}\right)^{3/2}
  \exp \left( - \frac{\bar{\bm{P}}^{2}}{2 \bar{M} k_{B} T}  \right), \\
 \label{local_equilibrium_bond_momentum_flexible_dumbbell}
 \psi_{\text{eq}}(\bm{p} | \bm{r})
 & = 
  \left(\frac{1}{2 \pi m k_{B} T}\right)^{3/2}
  \exp \left(- \frac{\bm{p}^{2}}{2 m k_{B} T} \right).
\end{align}
Then the stress tensor at the overdamped limit becomes
\begin{align}
 \label{effective_stress_tensor_dumbbell_total}
 \hat{\bm{\sigma}}^{\text{(OD)}}(\bar{\bm{R}},\bm{r})
 & = \hat{\bm{\sigma}}_{\text{CM}}^{\text{(OD)}}(\bar{\bm{R}})
 + \hat{\bm{\sigma}}_{\text{bond}}^{\text{(OD)}}(\bm{r}), \\
 \label{effective_stress_tensor_dumbbell_cm}
 \hat{\bm{\sigma}}_{\text{CM}}^{\text{(OD)}}(\bar{\bm{R}})
 & \equiv \int d\bar{\bm{P}} \,  \Psi_{\text{eq}}(\bar{\bm{P}} | \bar{\bm{R}}) \hat{\bm{\sigma}}_{\text{CM}}^{\text{(UD)}}(\bar{\bm{R}},\bar{\bm{P}})
 = - \frac{k_{B} T }{V} \bm{1}, \\
 \label{effective_stress_tensor_dumbbell_bond}
  \hat{\bm{\sigma}}_{\text{bond}}^{\text{(OD)}}(\bm{r})
 & \equiv \int d\bm{p} \,  \psi_{\text{eq}}(\bm{p} | \bm{r})
   \hat{\bm{\sigma}}_{\text{bond}}^{\text{(UD)}}(\bm{r},\bm{p})
 = \frac{1}{V} \left[ \frac{\partial \phi(\bm{r})}{\partial \bm{r}} \bm{r} - k_{B} T \bm{1} \right].
\end{align}
Except the trivial isotropic component, eq~\eqref{effective_stress_tensor_dumbbell_bond}
coincides to eq~\eqref{stress_tensor_dumbbell_bond_kramers}.

\subsection{Rigid Dumbbell}

We consider the rigid dumbbell model. In the rigid dumbbell model,
two particles are connected by a rigid bond instead of a
tethering potential. Since the dynamics of the center of mass is
independent of the bond, it is common for the flexible and rigid dumbbells.
What we need to consider is the dynamics of the bond and the stress tensor
by the bond.

If we naively
employ eqs~\eqref{effective_stress_tensor_dumbbell_total}-\eqref{effective_stress_tensor_dumbbell_bond}
for the stress tensor of the rigid dumbbell at the overdamped limit,
we cannot calculate the stress tensor.
The center of mass is independent of the bond vector and thus
eq~\eqref{effective_stress_tensor_dumbbell_cm} can be utilized without any modifications.
However, we cannot directly utilize eq~\eqref{effective_stress_tensor_dumbbell_bond}
because it contains the tethering potential $\phi(\bm{r})$.
If we employ a very stiff tethering potential with which the bond length becomes almost constant,
we can calculate the stress tensor as shown in Appendix~\ref{dumbbell_model_with_stiff_harmonic_dumbbell}.
But the stiff limit is not physically reasonable and the validity is not clear.
If we simply set $\phi(\bm{r}) = 0$ in eq~\eqref{effective_stress_tensor_dumbbell_bond},
we are led to conclude that the rigid dumbbell exhibits only the isotropic
stress. Clearly this is wrong. It is known that an anisotropic rigid body immersed in a fluid
generally exhibits an anisotropic stress tensor\cite{Doi-Edwards-book,Batchelor-1970}.
The origin of the anisotropic stress tensor is attributed to so-called
the Brownian potential\cite{Doi-Edwards-book} or the stress of the surrounding fluid\cite{Doi-Edwards-book,Batchelor-1970}.
However, in our coarse-grained description, we can not employ the Brownian potential
nor the effect of the surrounding fluid.

One way to avoid this difficulty is to go back to the underdamped system
and replace the potential force $- \partial \phi(\bm{r}) / \partial \bm{r}$
in eq~\eqref{stress_tensor_dumbbell_bond} by the constraint force acting on the bond.
If we replace the potential force by the constraint force $\bm{F}_{\text{constraint}}$,
we can construct the stress tensor at the underdamped system.
Under the rigid constraint, the local equilibrium distribution of the bond momentum 
can deviate from eq~\eqref{local_equilibrium_bond_momentum_flexible_dumbbell}.
Thus we should be careful when we take the overdamped limit.

We employ the expression for the centrifugal force to keep $r = |\bm{r}|$ constant
as the constraint force: $\bm{F}_{\text{constraint}} = - (\bm{p}^{2} / m r^{2}) \bm{r} $\cite{Landau-Lifshitz-book-mechanics}.
With this phenomenological replacement, we have
\begin{equation}
 \label{stress_tensor_dumbbell_bond_with_centrifugal_force}
 \hat{\bm{\sigma}}_{\text{bond}}^{(\text{UD})}(\bm{r},\bm{p})
 = \frac{1}{V} \left( \frac{\bm{p}^{2}}{m r^{2}} \bm{r} \bm{r} - \frac{\bm{p}{\bm{p}}}{m} \right).
\end{equation}
From the fact that the bond momentum $\bm{p}$ is perpendicular to $\bm{r}$ if $r$ is fixed,
$\bm{p}$ should be distributed on a plane which is perpendicular to $\bm{r}$.
Thus the local equilibrium distribution function for the bond momentum $\bm{p}$ depends on
the bond vector unlike eq~\eqref{local_equilibrium_bond_momentum_flexible_dumbbell}. The explicit form of the distribution 
function is:
\begin{equation}
 \label{equilibrium_distribution_bond_momentum_rigid}
 \psi_{\text{eq}}(\bm{p} | \bm{r})
  = \delta\left( \frac{\bm{r}\bm{r}}{r^{2}} \cdot \bm{p} \right)
  \frac{1}{2 \pi m k_{B} T}
  \exp \left(- \frac{\bm{p}^{2}}{2 m k_{B} T} \right).
\end{equation}
$\bm{r}\bm{r} / r^{2}$ works as the projection tensor which extracts the parallel component to $\bm{r}$ from a vector.
From eqs~\eqref{stress_tensor_dumbbell_bond_with_centrifugal_force} and \eqref{equilibrium_distribution_bond_momentum_rigid},
we have the following expression for the bond stress tensor at the overdamped limit:
\begin{equation}
 \label{effective_stress_tensor_dumbbell_bond_with_centrifugal_force}
  \hat{\bm{\sigma}}_{\text{bond}}^{\text{(OD)}}(\bm{r})
 = \frac{k_{B} T}{V} \left( \frac{3 \bm{r} \bm{r}}{r^{2}} - \bm{1} \right).
\end{equation}
(The detailed calculation is shown in Appendix~\ref{detailed_calculations}.)
Eq~\eqref{effective_stress_tensor_dumbbell_bond_with_centrifugal_force} states that
the anisotropic part of the stress tensor is proportional to $\bm{r}\bm{r}$.
This is consistent with the intuitive expectation, and thus we expect
that eq~\eqref{effective_stress_tensor_dumbbell_bond_with_centrifugal_force}
is reasonable.

Here we recall that
the stress tensor \eqref{stress_tensor_dumbbell_bond}
is originally constructed based on the energy change for the instantaneous
deformation by eq~\eqref{virtual_deformation_bond}. For the rigid
dumbbell, such a deformation is generally not allowed since the bond length
can be changed by eq~\eqref{virtual_deformation_bond}.
Also, the validity of the simple centrifugal force
to the instantaneous deformation is not clear.
Therefore, the
validity of eqs~\eqref{stress_tensor_dumbbell_bond_with_centrifugal_force}
and \eqref{effective_stress_tensor_dumbbell_bond_with_centrifugal_force}
is questionable from the viewpoint of the virtual work.
To be fair, we should mention that the same stress tensor can be obtained
without applying the virtual deformation\cite{Jongshaap-1987,Bird-Curtiss-1985}.
For example, the Irving-Kirkwood formalism\cite{Irving-Kirkwood-1950} can
give the same result without considering a virtual deformation.
But the Irving-Kirkwood formalism is not simple compared with 
the virtual work method. The virtual work method would be
preferred in some cases, if it gives the
correct stress tensor without a heuristic replacement.

To apply the virtual work method to the rigid dumbbell,
we should consider the dynamic equations for the rigid dumbbell carefully.
Without the flow field, the dynamic equations are well known.
The constraint force $\bm{F}_{\text{constraint}}(t)$ should be added to the dynamic equations, instead of the potential force\cite{Evans-Morriss-book}.
The constraint force is parallel to the bond vector, and thus we express it as
$\bm{F}_{\text{constraint}}(t) = \lambda(t) \bm{r}(t)$ with $\lambda(t)$ being
a time-depending scalar quantity.
By combining the effect of the SLLOD-type flow and the rigid constraint,
we have the following dynamic equations:
\begin{align}
 \label{dynamic_equation_r_sllod_original}
 \frac{d\bm{r}(t)}{dt} & = \frac{1}{m} \bm{p}(t)
  + \bm{\kappa}(t) \cdot \bm{r}(t), \\
 \label{dynamic_equation_p_sllod_original}
 \frac{d\bm{p}(t)}{dt} & = \lambda(t) \bm{r}(t)
 - \bm{\kappa}(t) \cdot \bm{p}(t)
  - \frac{\zeta}{m} \bm{p}(t) + \sqrt{2 \zeta k_{B} T} \bm{w}(t).
\end{align}
Here, $\lambda(t)$ can be interpreted as the Lagrange multiplier which is determined
to satisfy the constraint $|\bm{r}(t)| = b$ with $b$ being the bond length.
Similar dynamic equations have been employed to study the dynamics for more complex systems
such as alkanes\cite{Edberg-Evans-Morriss-1986,Edberg-Morriss-Evans-1986}. In numerical simulations, the dynamic equations
are discretized and the Lagrange multiplier $\lambda(t)$
is numerically determined at every time step.
In this work, we will calculate $\lambda(t)$ analytically.

\section{RESULTS}

\subsection{Dynamic Equations}

Although we expect that the dynamics of the rigid dumbbell
under flow can be described by eqs~\eqref{dynamic_equation_r_sllod_original} and \eqref{dynamic_equation_p_sllod_original},
it is apparently not clear how the rigid constraint $r(t) =  |\bm{r}(t)| = b$
is satisfied in eqs~\eqref{dynamic_equation_r_sllod_original} and \eqref{dynamic_equation_p_sllod_original}.
To apply the virtual deformation, eqs~\eqref{dynamic_equation_r_sllod_original} and \eqref{dynamic_equation_p_sllod_original}
do not seem to be convenient.
Therefore, first we attempt to rewrite eqs~\eqref{dynamic_equation_r_sllod_original}
and \eqref{dynamic_equation_p_sllod_original} and obtain dynamic equations which are suitable for our purpose.

From the rigid constraint, we have
\begin{equation}
 \label{holonomic_constraint_derivative}
 \frac{d}{dt} r^{2}(t) =
  2 \bm{r}(t) \cdot \frac{d \bm{r}(t)}{dt}  = 0.
\end{equation}
By combining eqs~\eqref{dynamic_equation_r_sllod_original} and \eqref{holonomic_constraint_derivative}, we find that
the right hand side of eq~\eqref{dynamic_equation_r_sllod_original} cannot
contain the component which is parallel to $\bm{r}(t)$:
\begin{equation}
 0 = \frac{1}{m} \bm{r}(t) \cdot \bm{p}(t) + 
  \bm{r}(t) \cdot \bm{\kappa}(t) \cdot \bm{r}(t).
\end{equation}
For convenience, we decompose the bond momentum $\bm{p}(t)$ into two components
which are parallel and
perpendicular to $\bm{r}(t)$: $\bm{p}(t) = \bm{p}_{\parallel}(t) + \bm{p}_{\perp}(t)$
with $\bm{p}_{\parallel}(t) \equiv [\bm{r}(t) \bm{r}(t) / r^{2}(t)] \cdot \bm{p}(t)$ and
$\bm{p}_{\perp}(t) \equiv [\bm{1} - \bm{r}(t) \bm{r}(t) / r^{2}(t)] \cdot \bm{p}(t)$.
Then we have
\begin{equation}
 \bm{p}_{\parallel}(t) 
  = - m \bm{\kappa}_{\parallel}(t) \cdot \bm{r}(t),
\end{equation}
with $\bm{\kappa}_{\parallel}(t) \equiv [\bm{r}(t) \bm{r}(t) / r^{2}(t)] \cdot \bm{\kappa}(t)$.
Now we can rewrite eq~\eqref{dynamic_equation_r_sllod_original} as
\begin{equation}
 \label{dynamic_equation_r_simple_sllod}
  \frac{d\bm{r}(t)}{dt}  = \frac{1}{m} \bm{p}_{\perp}(t)
  + \bm{\kappa}_{\perp}(t) \cdot \bm{r}(t).
\end{equation}
Here, $\bm{\kappa}_{\perp}(t) \equiv [\bm{1} - \bm{r}(t)\bm{r}(t) / r^{2}(t)] \cdot \bm{\kappa}(t)$
can be interpreted as the perpendicular component of the velocity gradient tensor.
The simple manipulation shown above tells us that 
the parallel component of the momentum $\bm{p}_{\parallel}(t)$ is not
an internal degree of freedom of the rigid dumbbell. This is not surprising
because the bond length $|\bm{r}(t)| = b$ is not an internal degree of
freedom, neither. We have only $4$ internal degrees of freedom for a rigid bond ($2$ for the
bond orientation and $2$ for the perpendicular bond momentum).
Eq~\eqref{dynamic_equation_r_simple_sllod} means that the bond vector effectively feels the
flow field which is perpendicular to $\bm{r}(t)$.

We want to rewrite the dynamic equation for the bond momentum $\bm{p}_{\perp}(t)$
(eq~\eqref{dynamic_equation_p_sllod_original})
in an explicit form. However, it is not that clear how we should manipulate eq~\eqref{dynamic_equation_p_sllod_original}.
We go back to the derivation of the SLLOD dynamic equation.
The SLLOD dynamic equation is designed to reproduce the following dynamic
equation with the external flow\cite{Evans-Morriss-1984,Baig-Edwards-Keffer-Cochran-2005}:
\begin{equation}
 \label{dynamic_equation_sllod_original_target}
 \frac{d^{2}\bm{r}(t)}{dt^{2}} = \frac{1}{m} \bm{F}(t) + \frac{d\bm{\kappa}(t)}{dt} \cdot \bm{r}(t)
 + O(\| \bm{\kappa}(t) \|^{2}),
\end{equation}
where $\bm{F}(t)$ is the force acting on the bond.
(In general, we have the second order term which is expressed as $O(\| \bm{\kappa}(t) \|^{2})$
in eq~\eqref{dynamic_equation_sllod_original_target}. We can remove the second
order term by employing so-called the {\it p}-SLLOD dynamic equations\cite{Baig-Edwards-Keffer-Cochran-2005}.
In this work, we simply ignore higher order terms.)
The force $\bm{F}(t)$ in eq~\eqref{dynamic_equation_sllod_original_target} can
be decomposed to the parallel and perpendicular component: $\bm{F}(t) = \bm{F}_{\parallel}(t) + \bm{F}_{\perp}(t)$.
The parallel component can be interpreted as the Lagrange multiplier $\bm{F}_{\parallel}(t) = \lambda(t) \bm{r}(t)$.
The perpendicular component contain the friction and noise terms: $\bm{F}_{\perp}(t) = - (\zeta / m) \bm{p}_{\perp}(t) + \sqrt{2 \zeta k_{B} T} \bm{w}_{\perp}(t)$,
where $\bm{w}_{\perp}(t)$ is the Gaussian white noise which is perpendicular to $\bm{r}(t)$.
(The parallel component of the friction and noise terms can be absorbed into $\lambda(t)$.)
We can rewrite eq~\eqref{dynamic_equation_sllod_original_target} as
\begin{equation}
 \label{dynamic_equation_sllod_original_target_modified}
 \begin{split}
  \frac{d^{2}\bm{r}(t)}{dt^{2}}
  & = \frac{1}{m} [\lambda(t) \bm{r}(t) +\bm{F}_{\perp}(t)]
  + \frac{d[\bm{\kappa}_{\perp}(t) + \bm{\kappa}_{\parallel}(t)]}{dt} \cdot \bm{r}(t) 
   + O(\| \bm{\kappa}(t) \|^{2}) \\  
  & = \frac{1}{m} 
  [\lambda'(t) \bm{r}(t) + \bm{F}_{\perp}(t)]
  + \frac{d\bm{\kappa}_{\perp}(t)}{dt} \cdot \bm{r}(t)
  + \frac{\tr \bm{\kappa}_{\parallel}(t)}{m}  \bm{p}_{\perp}(t)
  + O(\| \bm{\kappa}(t) \|^{2}). \\  
 \end{split}
\end{equation}
See Appendix~\ref{detailed_calculations} for the detailed calculations of eq~\eqref{dynamic_equation_sllod_original_target_modified}.
Here, $\tr$ represents the trace (for a second order tensor $\bm{A}$, $\tr \bm{A} \equiv \sum_{\alpha} A_{\alpha \alpha}$).
We have absorbed all the terms which are proportional to $\bm{r}(t)$ into $\lambda'(t)$. ($\lambda'(t)$ contains
the contribution from $[d\bm{\kappa}(t) / dt] \cdot \bm{r}(t)$ and is
generally different from $\lambda(t)$.)
On the other hand, by taking the time derivative of eq~\eqref{dynamic_equation_r_simple_sllod},
we have the following equation:
\begin{equation}
 \label{dynamic_equation_r_simple_sllod_derivative}
   \frac{d^{2}\bm{r}(t)}{dt^{2}} 
    = 
   \frac{1}{m} \frac{d\bm{p}_{\perp}(t)}{dt} 
   + \frac{d\bm{\kappa}_{\perp}(t)}{dt} \cdot \bm{r}(t) 
 + \frac{1}{m} \bm{\kappa}_{\perp}(t) \cdot \bm{p}_{\perp}(t)
   + \bm{\kappa}_{\perp}(t) \cdot \bm{\kappa}_{\perp}(t) \cdot \bm{r}(t) .
\end{equation}
By comparing eqs~\eqref{dynamic_equation_sllod_original_target_modified}
and \eqref{dynamic_equation_r_simple_sllod_derivative},
we find that the dynamic equation for $\bm{p}_{\perp}(t)$ should be given as
\begin{equation}
 \frac{d\bm{p}_{\perp}(t)}{dt}
  = \lambda'(t) \bm{r}(t) + \bm{F}_{\perp}(t)
  - [\bm{\kappa}_{\perp}(t) 
  - \bm{1} \tr \bm{\kappa}_{\parallel}(t)] \cdot \bm{p}_{\perp}(t) .
\end{equation}
The Lagrange multiplier $\lambda'(t)$ can be determined from the constraint $\bm{r}(t) \cdot \bm{p}_{\perp}(t) = 0$.
By taking the time derivative of this constraint, we have
\begin{equation}
 \begin{split}
  0
  & =  \bm{r}(t) \cdot \frac{d\bm{p}_{\perp}(t)}{dt} + \bm{p}_{\perp}(t) \cdot \frac{d\bm{r}(t)}{dt} \\
 & =   \lambda'(t) r^{2}(t) + \frac{1}{m} \bm{p}_{\perp}^{2}(t) + \bm{p}_{\perp}(t) \cdot \bm{\kappa}_{\perp}(t) \cdot \bm{r}(t),
 \end{split}
\end{equation}
and the explicit form of $\lambda'(t)$ is given as
\begin{equation}
 \label{simple_sllod_lagrange_multiplier}
 \lambda'(t) = - \frac{ \bm{p}_{\perp}^{2}(t)}{m r^{2}(t)}
  - \frac{\bm{r}(t) \cdot \bm{\kappa}^{\mathrm{T}}(t)  \cdot \bm{p}_{\perp}(t)}{r^{2}(t)}.
\end{equation}
Finally we have the explicit form of the dynamic equation for the bond momentum $\bm{p}_{\perp}(t)$:
\begin{equation}
 \label{dynamic_equation_p_simple_sllod}
\begin{split}
  \frac{d\bm{p}_{\perp}(t)}{dt} & = 
  - \frac{\bm{p}_{\perp}^{2}(t)}{m r^{2}(t)} \bm{r}(t)
 - \frac{\zeta}{m} \bm{p}_{\perp}(t) + \sqrt{2 \zeta k_{B} T} \bm{w}_{\perp}(t) \\
 & \qquad - [\bm{\kappa}_{\perp}(t) + (\bm{\kappa}^{\mathrm{T}})_{\parallel}(t)
 - \bm{1} \mathrm{Tr} \, \bm{\kappa}_{\parallel}(t)] \cdot \bm{p}_{\perp}(t),
\end{split}
\end{equation}
with $(\bm{\kappa}^{\mathrm{T}})_{\parallel}(t) \equiv [\bm{r}(t) \bm{r}(t) / r^{2}(t) ] \cdot \bm{\kappa}^{\mathrm{T}}(t)$.
The noise $\bm{w}_{\perp}(t)$ satisfies the following relations:
\begin{equation}
 \label{fluctuation_dissipation_simple_sllod}
  \langle \bm{w}_{\perp}(t) \rangle = 0, \qquad
  \langle \bm{w}_{\perp}(t) \bm{w}_{\perp}(t') \rangle = \left[\bm{1} - \frac{\bm{r}(t)\bm{r}(t)}{r^{2}(t)} \right]\delta(t - t').
\end{equation}
Eqs~\eqref{dynamic_equation_r_simple_sllod} and \eqref{dynamic_equation_p_simple_sllod}
are the dynamic equations for the bond vector of the rigid dumbbell under flow.

\subsection{Virtual Work Method}

We can apply the instantaneous virtual deformation to the system
by utilizing eqs~\eqref{dynamic_equation_r_simple_sllod} and \eqref{dynamic_equation_p_simple_sllod}.
As before, we set $\bm{\kappa}(t) = \bm{E} \delta(t)$ and integrate
the dynamic equations from $t = -0$ to $t = +0$.
The contribution of the center of mass is common for the flexible and rigid dumbbells.
The stress tensor by the center of mass is given as
eq~\eqref{stress_tensor_dumbbell_cm} (for the underdamped system) or
eq~\eqref{effective_stress_tensor_dumbbell_cm} (for the overdamped system).
Thus we consider only the contribution of the bond vector in what follows.

The parallel component of the bond momentum is zero: $\bm{p}_{\parallel}' = \bm{p}_{\parallel} = 0$.
By integrating eqs~\eqref{dynamic_equation_r_simple_sllod} and \eqref{dynamic_equation_p_simple_sllod},
the bond vector and the perpendicular component of the bond moment change as
\begin{align}
 \label{impulse_deformation_q}
 \bm{r}'
 & = \bm{r} + \left( \bm{1} - \frac{\bm{r}\bm{r}}{r^{2}} \right) \cdot \bm{E} \cdot \bm{r} + O(\| \bm{E} \|^{2}), \\
 \label{impulse_deformation_p}
 \begin{split}
  \bm{p}'_{\perp}
  & = \bm{p}_{\perp}
  - \left( \bm{1} - \frac{\bm{r}\bm{r}}{r^{2}} \right) \cdot \bm{E} \cdot \bm{p}_{\perp}
  + \frac{\bm{r} \cdot \bm{E} \cdot \bm{r}}{r^{2}} \bm{p}_{\perp}
      - \frac{ \bm{p}_{\perp} \cdot \bm{E} \cdot \bm{r}}{r^{2}} \bm{r} + O(\|\bm{E}\|^{2}).
 \end{split}
\end{align}
Here, we have utilized the identity $(\bm{E}^{\mathrm{T}})_{\parallel} \cdot \bm{p}_{\perp}
= \bm{r} (\bm{r} \cdot \bm{E}^{\mathrm{T}} \cdot \bm{p}_{\perp}) / r^{2}
= (\bm{p}_{\perp} \cdot \bm{E} \cdot \bm{r} / r^{2}) \bm{r}$
to derive eq~\eqref{impulse_deformation_p}.
In eqs~\eqref{impulse_deformation_q} and \eqref{impulse_deformation_p},
the higher order terms in $\bm{E}$ are not explicitly shown because their contribution to the energy change is negligibly small.
(For our purpose, only the first order terms are required.)
The kinetic energy of the bond is changed as
\begin{equation}
 \frac{(\bm{p}'_{\perp})^{2}}{2 m} -  \frac{\bm{p}_{\perp}^{2}}{2 m}
  = \left[ - \frac{\bm{p}_{\perp}\bm{p}_{\perp}}{m}
  +   \frac{\bm{p}^{2}_{\perp}}{m} \frac{\bm{r} \bm{r}}{r^{2}} \right] : \bm{E} 
+ O(\|\bm{E}\|^{2}).
\end{equation}
Then we have the stress tensor for the single bond as
\begin{equation}
 \label{stress_tensor_dumbbell_bond_with_constraint}
 \hat{\bm{\sigma}}_{\text{bond}}^{(\text{UD})}(\bm{r},\bm{p}_{\perp}) = \frac{1}{V} \left[ - \frac{\bm{p}_{\perp}\bm{p}_{\perp}}{m}
  +   \frac{\bm{p}^{2}_{\perp}}{m} \frac{\bm{r} \bm{r}}{r^{2}} \right].
\end{equation}
Eq~\eqref{stress_tensor_dumbbell_bond_with_constraint} coincides to
eq~\eqref{stress_tensor_dumbbell_bond_with_centrifugal_force}.
If we take the local equilibrium average over the bond momentum, we have
eq~\eqref{effective_stress_tensor_dumbbell_bond_with_centrifugal_force}.

Therefore, our method gives
the same result with the simple heuristic method in which the potential force is
replaced by the constraint force.
Here we emphasize that we did not employ such a heuristic replacement.
What we did is just to rewrite the SLLOD dynamic equations and applied
the virtual deformation to the system which does not violate the rigid constraint.
According to our derivation,
the stress tensor by the bond (eq~\eqref{stress_tensor_dumbbell_bond_with_constraint})
purely comes from the kinetic energy.

We expect that our method can be applied to other systems with
different thermostats. For example, we will be able to construct the stress
tensor of a rigid dumbbell driven by the SLLOD dynamic equations with the
Nose-Hoover thermostat, in the same way as this work. What important in our formalism is the explicit
expressions for the advection terms in the dynamic equations (such as the last term in the right hand side of eq~\eqref{dynamic_equation_p_simple_sllod}),
and they will not be affected by the specific choice of the thermostat.
The resulting expression of the stress tensor does not depend on whether
the system is in equilibrium or not.
Therefore, our method will be useful to analyze the systems where the friction coefficients
and the Brownian force intencities are modulated by the imposed flows.\cite{Watanabe-Matsumiya-Sato-2020}

This is in contrast to the virtual work method for the distribution function.
To derive the stress tensor, sometimes the Fokker-Planck equation for the
distribution function is utilized instead of the Langevin equation.
In this approach, we apply the virtual deformation to the distribution function
and calculate the energy change. This energy change contains the change of the
Brownian potential arisen from the random noise\cite{Doi-Edwards-book}.
Therefore, naively we expect that the modulation of the friction coefficient and the Brownian
force may affect the expression of the stress tensor. However, the Brownian
potential is not a potential for a specific dumbbell, but for the ensemble
of dumbbells. The interpretation of the energy change is thus not clear.
In our method, we do not consider the ensemble and there is no such ambiguity.

\subsection{Relaxation Modulus at Overdamped Limit}

We consider the situation where the momentum relaxation is sufficiently fast.
Under such a situation, we take the overdamped limit and eliminate the
degree of freedom of the momenta. By setting $d\bm{p}(t) /dt = 0$ and
$\bm{\kappa}(t) = 0$ in eq~\eqref{dynamic_equation_p_sllod_original},
we can eliminate $\bm{p}_{\perp}(t)$ from eq~\eqref{dynamic_equation_r_sllod_original}.
We can rewrite eq~\eqref{dynamic_equation_r_sllod_original} as
\begin{equation}
 \label{dynamic_equation_r_sllod_overdamped}
  \frac{d\bm{r}(t)}{dt} = - \lambda''(t) \bm{r}(t)
  + \sqrt{\frac{2 k_{B} T}{\zeta}} \bm{w}_{\perp}(t)
  + \bm{\kappa}_{\perp}(t) \cdot \bm{r}(t),
\end{equation}
where $\lambda''(t)$ is the Lagrange multiplier, and all the terms which are 
parallel to $\bm{r}(t)$ are absorbed into the Lagrange multiplier.
$\lambda''(t)$ can be determined from the constraint $d r^{2}(t) / dt = 0$.
By using the Ito formula\cite{vanKampen-book,Gardiner-book} together with
eqs~\eqref{fluctuation_dissipation_simple_sllod},
\eqref{dynamic_equation_r_sllod_overdamped} and $\partial^{2} (r^{2}) / \partial \bm{r} \partial \bm{r} = 2 \bm{1}$, we have
\begin{equation}
\begin{split}
 0 & = 2 \bm{r}(t) \cdot \frac{d\bm{r}(t)}{dt}
 + \frac{1}{2} \frac{k_{B} T}{\zeta} \left[\bm{1} - \frac{\bm{r}(t)\bm{r}(t)}{r^{2}(t)}\right] : 2 \bm{1} \\
  & = 2 \bm{r}(t) \cdot \left[ \lambda''(t) \bm{r}(t)
  + \sqrt{\frac{2 k_{B} T}{\zeta}} \bm{w}_{\perp}(t)
  + \bm{\kappa}_{\perp}(t) \cdot \bm{r}(t) \right]
 + \frac{k_{B} T}{\zeta} \tr \left[\bm{1} - \frac{\bm{r}(t)\bm{r}(t)}{r^{2}(t)}\right] \\
 & = - 2 \lambda''(t) r^{2}(t) + \frac{2 k_{B} T}{\zeta}.
\end{split}
\end{equation}
Then we have $\lambda''(t) = k_{B} T / \zeta r^{2}(t)$ and
the overdamped dynamic equation for the bond becomes
\begin{equation}
 \label{dynamic_equation_r_sllod_overdamped_modified}
 \frac{d\bm{r}(t)}{dt} = - \frac{k_{B} T}{\zeta} \frac{\bm{r}(t)}{r^{2}(t)} 
  + \sqrt{\frac{2 k_{B} T}{\zeta}} \bm{w}_{\perp}(t)
  + \bm{\kappa}_{\perp}(t) \cdot \bm{r}(t).
\end{equation}
As the case of the underdamped dynamics, only the perpendicular component
of the velocity gradient tensor is applied to the bond vector.

We can calculate
rheological properties of a rigid dumbbell model at the overdamped limit by combining
eqs~\eqref{effective_stress_tensor_dumbbell_bond_with_centrifugal_force}
and \eqref{dynamic_equation_r_sllod_overdamped_modified}.
To demonstrate that we can reasonably describe the rheological
properties of a rigid dumbbell,
here we derive the linear response formula and
calculate the shear relaxation modulus.
The probability distribution function for the bond vector $\psi(\bm{r},t)$ obeys the following Fokker-Planck equation:
\begin{equation}
 \label{fokker_planck_equation_overdamped}
 \frac{\partial \psi(\bm{r},t)}{\partial t}
  = [\mathcal{L}_{\text{eq}}
  + \Delta\mathcal{L}(t) ] \psi(\bm{r},t) ,
\end{equation}
with the Fokker-Planck operators defined as
\begin{align}
 \label{fokker_planck_operator_equilibrium}
 \mathcal{L}_{\text{eq}} \psi(\bm{r}) 
 & \equiv \frac{k_{B} T}{\zeta} \frac{\partial}{\partial \bm{r}} \cdot 
  \left[ \left(\bm{1} - \frac{\bm{r}\bm{r}}{r^{2}} \right) \cdot 
 \frac{\partial \psi(\bm{r},t)}{\partial \bm{r}}
  \right], \\
 \Delta \mathcal{L}(t) \psi(\bm{r}) 
 & \equiv - \frac{\partial}{\partial \bm{r}} \cdot 
  \left[ \left(\bm{1} - \frac{\bm{r}\bm{r}}{r^{2}} \right) \cdot 
    \bm{\kappa}(t) \cdot \bm{r} \psi(\bm{r},t)
  \right].
\end{align}
$\mathcal{L}_{\text{eq}}$ describes the rotational diffusion in equilibrium
whereas $\Delta \mathcal{L}(t)$ describes the advection by the applied flow.

We can construct the linear response theory by treating the applied
velocity gradient as the perturbation. We split the distribution function
into the equilibrium and perturbation parts as $\psi(\bm{r},t) = \psi_{\text{eq}}(\bm{r}) + \Delta \psi(\bm{r},t)$,
with $\psi_{\text{eq}}(\bm{r})$ being the equilibrium distribution function of the bond vector
and $\Delta \psi(\bm{r},t)$ being the time-dependent perturbation part of the distribution function.
In equilibrium, the bond vector is uniformly distributed on a sphere.
The equilibrium distribution function depends only on $r = |\bm{r}|$:
\begin{equation}
 \label{bond_vector_equilibrium_distribution}
 \psi_{\text{eq}}(\bm{r}) =  \psi_{\text{eq}}(r) = \frac{1}{4 \pi b^{2}} \delta(r - b).
\end{equation}
We interpret $\Delta\psi(\bm{r},t)$ and $\Delta\mathcal{L}(t)$ as perturbations.
At the first order
in the perturbation, the perturbation part of the distribution function satisfies the following
equation:
\begin{equation}
 \label{fokker_planck_equation_overdamped_perturbation}
 \frac{\partial \Delta \psi(\bm{r},t)}{\partial t}
  = \Delta \mathcal{L}(t) \psi_{\text{eq}}(\bm{r}) 
  + \mathcal{L}_{\text{eq}} \Delta \psi(\bm{r},t) .
\end{equation}
The solution of eq~\eqref{fokker_planck_equation_overdamped_perturbation} is
\begin{equation}
 \label{fokker_planck_equation_overdamped_perturbation_solution}
\begin{split}
  \Delta \psi(\bm{r},t)
 & = \int_{-\infty}^{t} dt' \, e^{ (t - t') \mathcal{L}_{\text{eq}}} \Delta \mathcal{L}(t) \psi_{\text{eq}}(\bm{r}) \\
 & = \frac{V}{k_{B} T} \int_{-\infty}^{t} dt' \, e^{- (t - t') \mathcal{L}_{\text{eq}}}  \hat{\bm{\sigma}}_{\text{bond}}^{\text{(OD)}}(\bm{r}) :
    \bm{\kappa}(t) \psi_{\text{eq}}(\bm{r}) . 
\end{split}
\end{equation}
In the calculation of eq~\eqref{fokker_planck_equation_overdamped_perturbation_solution}, we have utilized the following relation:
\begin{equation}
 \begin{split}
  \Delta \mathcal{L}(t) \psi_{\text{eq}}(\bm{r})
  & = \left[  \frac{2 \bm{r}}{r^{2}} \cdot 
    \bm{\kappa}(t) \cdot \bm{r}
- \tr 
  \left[ \left(\bm{1} - \frac{\bm{r}\bm{r}}{r^{2}} \right) \cdot 
    \bm{\kappa}(t) 
  \right] \right] \psi_{\text{eq}}(\bm{r})
  \\
  & = \frac{V}{k_{B} T} \hat{\bm{\sigma}}_{\text{bond}}^{\text{(OD)}}(\bm{r}) : \bm{\kappa}(t) \psi_{\text{eq}}(\bm{r}).
 \end{split}
\end{equation}
The second order or higher order perturbation will be negligible if the
applied flow is sufficiently weak.
Then we can calculate the (ensemble) average stress tensor at time $t$ under the applied velocity
gradient history with eq~\eqref{fokker_planck_equation_overdamped_perturbation_solution}:
\begin{equation}
 \label{average_stress_tensor_overdamped_perturbation}
\begin{split}
  \bm{\sigma}_{\text{bond}}^{\text{(OD)}}(t)
  & = \int d\bm{r} \, \hat{\bm{\sigma}}_{\text{bond}}^{\text{(OD)}}(\bm{r})
 [\psi_{\text{eq}}(\bm{r}) + \Delta \psi(\bm{r},t)] \\
  & = \frac{V}{k_{B} T} \int_{-\infty}^{t} dt' \int d\bm{r} \, \hat{\bm{\sigma}}_{\text{bond}}^{\text{(OD)}}(\bm{r}) 
 e^{ (t - t') \mathcal{L}_{\text{eq}}}  \hat{\bm{\sigma}}_{\text{bond}}^{\text{(OD)}}(\bm{r}) :
    \bm{\kappa}(t) \psi_{\text{eq}}(\bm{r}) \\
  & = \frac{V}{k_{B} T} \int_{-\infty}^{t} dt' \int d\bm{r} \, \psi_{\text{eq}}(\bm{r})
 \left[ e^{ (t - t') \mathcal{L}_{\text{eq}}^{\dagger}} \hat{\bm{\sigma}}_{\text{bond}}^{\text{(OD)}}(\bm{r}) \right]
 \hat{\bm{\sigma}}_{\text{bond}}^{\text{(OD)}}(\bm{r}) :
    \bm{\kappa}(t)  .
\end{split}
\end{equation}
Here, $\mathcal{L}_{\text{eq}}^{\dagger}$ is the adjoint Fokker-Planck operator of $\mathcal{L}_{\text{eq}}$,
and $e^{(t - t') \mathcal{L}_{\text{eq}}^{\dagger}}$ works as the time-shift operator.

If the perturbation is sufficiently weak, we can assume that
the system behaves as the linear viscoelastic material.
The average stress tensor can be related to the velocity gradient history
by using the relaxation modulus tensor $\bm{\Lambda}(t)$:
\begin{equation}
 \label{average_stress_tensor_macroscopic_relation}
  \bm{\sigma}^{\text{(OD)}}(t)
  = - P_{\text{eq}} \bm{1} + \int_{-\infty}^{t} dt' \, \bm{\Lambda}(t - t') : \bm{\kappa}(t'),
\end{equation}
with $P_{\text{eq}} \equiv k_{B} T / V$ being the equilibrium pressure.
By comparing eqs~\eqref{average_stress_tensor_overdamped_perturbation}
and \eqref{average_stress_tensor_macroscopic_relation},
we have the linear response formula for the relaxation modulus tensor of a single rigid dumbbell:
\begin{equation}
 \label{relaxation_modulus_tensor_overdamped}
 \bm{\Lambda}(t)
  = \frac{V}{k_{B} T} \left\langle \hat{\bm{\sigma}}_{\text{bond}}^{\text{(OD)}}(\bm{r},t)
    \hat{\bm{\sigma}}_{\text{bond}}^{\text{(OD)}}(\bm{r}) \right\rangle_{\text{eq}},
\end{equation}
where $\langle \dots \rangle_{\text{eq}}$ represents the equilibrium statistical average
and $\hat{\bm{\sigma}}_{\text{bond}}^{\text{(OD)}}(\bm{r},t) \equiv e^{t \mathcal{L}^{\dagger}_{\text{eq}}} \hat{\bm{\sigma}}(\bm{r})$
is the time-shifted stress tensor.
Eq~\eqref{relaxation_modulus_tensor_overdamped} is nothing but the Green-Kubo
relation.
Judging from the fact that eqs~\eqref{effective_stress_tensor_dumbbell_bond_with_centrifugal_force}
and \eqref{dynamic_equation_r_sllod_overdamped_modified}
reproduce the Green-Kubo relation, we conclude that they reasonably describe the dynamics and rheology
of the rigid dumbbell.

The shear relaxation modulus of a dilute suspension which consist of $N$ rigid dumbbells is
\begin{equation}
 \label{shear_relaxation_modulus_dumbbells_overdamped}
 G(t) = N \Lambda_{xyxy}(t) 
  = \frac{9 \nu k_{B} T }{b^{4}}
  \left\langle  r_{x}(t) r_{y}(t) r_{x}(0) r_{y}(0)\right\rangle_{\text{eq}}.
\end{equation}
where $\nu = N / V$ is the dumbbell number density.
The correlation function for the bond vector in equilibrium can be
evaluated analytically\cite{Doi-Edwards-book}.
We have the following explicit expression for the relaxation modulus:
\begin{equation}
 \label{shear_relaxation_modulus_dumbbells_overdamped_final}
  G(t) = \frac{3}{5}  \nu k_{B} T 
  \exp\left( - \frac{6  k_{B} T}{\zeta b^{2}} t \right).
\end{equation}
See Appendix~\ref{detailed_calculations} for the detailed calculations.
Eq~\eqref{shear_relaxation_modulus_dumbbells_overdamped_final} coincides to
the relaxation modulus calculated by the kinetic theory, except the instantaneous contribution
which is proportional to $\delta(t)$ \cite{Bird-Armstrong-Hassager-book,Bird-Curtiss-Armstrong-Hassager-book}.
This result supports the validity of eq~\eqref{relaxation_modulus_tensor_overdamped},
and also the validity of eqs~\eqref{effective_stress_tensor_dumbbell_bond_with_centrifugal_force}
and \eqref{dynamic_equation_r_sllod_overdamped_modified}.

\section{CONCLUSIONS}

We calculated the stress tensor of the rigid dumbbell based on the
virtual work method. To apply the virtual deformation and calculate
the energy change, we considered the underdamped SLLOD-type dynamic equations
for the bond vector. Due to the rigid constraint, the bond vector cannot
affinely move following the applied velocity gradient tensor.
We rewrote the dynamic equations in which the bond vector and the bond momentum
are driven by the effective velocity gradient tensor
(eqs~\eqref{dynamic_equation_r_simple_sllod} and \eqref{dynamic_equation_p_simple_sllod}).
We derived the explicit expression for the Lagrange multiplier in the dynamic equation,
and the dynamic equations are expressed in an explicit form.
With the thus derived dynamic equations, we applied the virtual deformation
to the system. Then, from the change of the kinetic energy before and
after the small impulsive deformation, we obtained the stress tensor of
the single bond (eq~\eqref{stress_tensor_dumbbell_bond_with_constraint}).
Finally we considered the overdamped limit and derived the linear response
formula for the single bond (eq~\eqref{relaxation_modulus_tensor_overdamped}).
We showed that the shear relaxation modulus calculated by the linear response formula
is consistent with the modulus calculated by the kinetic theory,
except the short-time scale delta function type contribution.

The results in this work will be informative to study other systems with 
rigid constraints. For example, stress of the freely jointed chain where
the beads are connected by rigid bonds will be handled in a similar way to
our method (although the analytic expressions will be too complicated).
Our results suggest that the virtual work method can be applied
to other systems with strong constraints. The hard sphere potentials can
be interpreted as constraint rather than an interaction potential.
It would be interesting to formulate the dynamic equations and the
stress tensor of hard sphere systems in a similar way to this work.

\section*{ACKNOWLEDGMENT}

This work was supported by JST, PRESTO Grant Number JPMJPR1992, Japan,
Grant-in-Aid (KAKENHI) for Scientific Research Grant B No.~JP19H01861,
and Grant-in-Aid (KAKENHI) for Transformative Research Areas B JP20H05736.

\appendix

\section*{APPENDIX}
\label{appendix}

\section{Dumbbell Model with Stiff Harmonic Potential}
\label{dumbbell_model_with_stiff_harmonic_dumbbell}

The rigid constraint may be approximated by a ``stiff'' tethering potential.
In this appendix, we consider the case where the tethering potential is given
as the following harmonic potential:
\begin{equation}
 \label{tethering_potential_stiff_harmonic}
 \phi(\bm{r}) = \frac{1}{2} K (r  - b)^{2}.
\end{equation}
Here, $r = |\bm{r}|$ is the bond length, $K$ is the spring constant and $b$ is the natural bond length.
We consider the case where $K$ is sufficiently large: $K \gg k_{B} T$.
Under such a condition, the bond length can only slightly fluctuate
around the average value $b$.
If we take the limit of $K \to \infty$, the bond length fluctuation
will approach to zero and the bond length will
be constant ($r \to b$).

If we naively assume the overdamped limit,
from eqs~\eqref{stress_tensor_dumbbell_bond_kramers} and
\eqref{tethering_potential_stiff_harmonic}, we have
\begin{equation}
 \label{stress_tensor_dumbbell_bond_kramers_stiff_harmonic}
 \sigma_{\text{bond}}^{(\text{OD})}(\bm{r})
  = \frac{1}{V} K r (r - b) \frac{\bm{r}\bm{r}}{r^{2}}.
\end{equation}
The equilibrium distribution of the bond length $r$ can be approximately
expressed as
\begin{equation}
 \psi_{\text{eq}}(r) \approx \sqrt{\frac{K}{2 \pi k_{B} T}}
  \exp\left[ - \frac{K}{2 k_{B} T} (r - b)^{2} \right].
\end{equation}
Then, by taking the partial average of eq~\eqref{stress_tensor_dumbbell_bond_kramers_stiff_harmonic}
over the bond length, we have
\begin{equation}
 \label{stress_tensor_dumbbell_bond_kramers_stiff_harmonic_average}
  \int dr \, \sigma_{\text{bond}}^{(\text{OD})}(\bm{r}) \psi_{\text{eq}}(r)
  = \frac{k_{B} T}{V} \frac{\bm{r}\bm{r}}{r^{2}}. 
\end{equation}
Eq~\eqref{stress_tensor_dumbbell_bond_kramers_stiff_harmonic_average} does not
coincide to the correct expression (eq~\eqref{effective_stress_tensor_dumbbell_bond_with_centrifugal_force}).

The reason why we do not have the correct stress tensor with
eqs~\eqref{stress_tensor_dumbbell_bond_kramers} and
\eqref{tethering_potential_stiff_harmonic} is rather simple.
The characteristic relaxation time of the bond length becomes very short
when $K \gg k_{B} T$. This means that we cannot simply take the overdamped
limit. Thus we should consider the underdamped system as in the main text.

Even if we consider the underdamped system, the situation is not that clear.
If the spring constant is very large, the quantum effect becomes non-negligible.
This situation is similar to thermodynamic properties of the diatomic gases\cite{Landau-Lifshitz-book-statphys}.
The specific heat of the diatomic gas consists of several different contributions.
The vibrational motion of the bond contributes to the specific heat only at
the relatively high temperature ($T \gg \hbar \omega$, with $\hbar$ being the reduced Planck constant
and $\omega$ being the characteristic angular frequency of the vibration).
If the temperature is relatively low ($T \ll \hbar \omega$),
the specific heat coincides to that of rigid diatomic gases. This is because
the vibration modes cannot take the excited states and the vibrational mode becomes
essentially frozen. The flexible dumbbell with a sufficiently stiff tethering
potential has very large characteristic angular frequency ($\omega \to \infty$ at the limit of $K \to \infty$), and thus it
will behave as the rigid dumbbell.

\section{Detailed Calculations}
\label{detailed_calculations}

In this appendix, we show detailed calculations for some relations used
in the main text. We show the calculations for the stress tensor of a single rigid dumbbell at the
over damped limit, eq~\eqref{effective_stress_tensor_dumbbell_bond_with_centrifugal_force}.
To calculate the overdamped limit, we need to calculate the partial average over
the bond momentum $\bm{p}$. It is convenient to decompose the bond momentum
into the parallel and perpendicular components: $\bm{p} = \bm{p}_{\parallel} + \bm{p}_{\perp}$
with $\bm{p}_{\parallel} = (\bm{r} \bm{r} / r^{2}) \cdot \bm{p}$ and
 $\bm{p}_{\perp} = (\bm{1} - \bm{r} \bm{r} / r^{2}) \cdot \bm{p}$.
Eq~\eqref{equilibrium_distribution_bond_momentum_rigid} can be rewritten as
\begin{equation}
 \label{equilibrium_distribution_bond_momentum_rigid_modified}
 \psi_{\text{eq}}(\bm{p} | \bm{r})
  = \delta\left( \bm{p}_{\parallel} \right)
  \frac{1}{2 \pi m k_{B} T}
  \exp \left(- \frac{\bm{p}_{\perp}^{2}}{2 m k_{B} T} \right).
\end{equation}
Then stress tensor at the overdamped limit can be calculated as
\begin{equation}
 \label{stress_tensor_dumbbell_bond_with_centrifugal_force_overdamped_limit}
  \begin{split}
   \hat{\bm{\sigma}}_{\text{bond}}^{\text{(OD)}}(\bm{r})
   & = \int d\bm{p} \, \psi_{\text{eq}}(\bm{p} | \bm{r}) \hat{\bm{\sigma}}_{\text{bond}}^{(\text{UD})}(\bm{r},\bm{p}) \\
   & =   \frac{1}{2 \pi m k_{B} T V}
\int d\bm{p}_{\parallel} d\bm{p}_{\perp}  \, 
   \delta\left( \bm{p}_{\parallel} \right)
  \exp \left(- \frac{\bm{p}_{\perp}^{2}}{2 m k_{B} T} \right) \\
   & \qquad \times 
   \left[ \frac{\bm{p}_{\perp}^{2} + \bm{p}_{\parallel}^{2}}{m r^{2}} \bm{r} \bm{r} 
   - \frac{(\bm{p}_{\parallel} + \bm{p}_{\perp}) (\bm{p}_{\parallel} + \bm{p}_{\perp})}{m} \right] \\
   & =   \frac{1}{2 \pi m k_{B} T V}
\int d\bm{p}_{\perp}  \, 
  \exp \left(- \frac{\bm{p}_{\perp}^{2}}{2 m k_{B} T} \right) 
   \left( \frac{\bm{p}_{\perp}^{2}}{m r^{2}} \bm{r} \bm{r} 
   - \frac{\bm{p}_{\perp} \bm{p}_{\perp}}{m} \right).
  \end{split}
\end{equation}
Here, $\bm{p}_{\perp}$ is distributed on
the two dimensional plane which is perpendicular to $\bm{r}$.
We express it as $\bm{p}_{\perp} = \xi \bm{e}_{1} + \eta \bm{e}_{2}$ with $\bm{e}_{1}$
and $\bm{e}_{2}$ being two orthogonal unit vectors which are perpendicular to $\bm{r}$.
($|\bm{e}_{1}| = |\bm{e}_{2}| = 1$, $\bm{e}_{1} \cdot \bm{e}_{2} = 0$, and $\bm{e}_{1} \cdot \bm{r} = \bm{e}_{2} \cdot \bm{r} = 0$.)
We have the following relation:
\begin{equation}
 \label{second_order_moment_gaussian_two_dimensional_plane}
 \begin{split}
  & \frac{1}{2 \pi m k_{B} T}
  \int d\bm{p}_{\perp}  \, 
  \exp \left(- \frac{\bm{p}_{\perp}^{2}}{2 m k_{B} T} \right) 
  \bm{p}_{\perp} \bm{p}_{\perp} \\
  & = \frac{1}{2 \pi m k_{B} T}
  \int d\xi d\eta  \, 
  \exp \left(- \frac{\xi^{2} + \eta^{2}}{2 m k_{B} T} \right) \\
  & \qquad \times [\xi^{2} \bm{e}_{1}\bm{e}_{1} 
  + \xi\eta (\bm{e}_{1}\bm{e}_{2} + \bm{e}_{2}\bm{e}_{1})
  + \eta^{2} \bm{e}_{2}\bm{e}_{2}] \\
  & = \bm{e}_{1}\bm{e}_{1} + \bm{e}_{2}\bm{e}_{2}  = \bm{1} - \frac{\bm{r}\bm{r}}{r^{2}}.
 \end{split}
\end{equation}
In the last line of eq~\eqref{second_order_moment_gaussian_two_dimensional_plane},
we utilized the fact that the second order tensor $\bm{e}_{1}\bm{e}_{1} + \bm{e}_{2}\bm{e}_{2}$
is the unit tensor on the two dimensional plane which is perpendicular to $\bm{r}$.
By combining eqs~\eqref{stress_tensor_dumbbell_bond_with_centrifugal_force_overdamped_limit} 
and \eqref{second_order_moment_gaussian_two_dimensional_plane}, we have eq~\eqref{effective_stress_tensor_dumbbell_bond_with_centrifugal_force}
in the main text.

We show the calculations for \eqref{dynamic_equation_sllod_original_target_modified}.
The term $[d\bm{\kappa}_{\parallel}(t) / dt] \cdot \bm{r}(t)$ in eq~\eqref{dynamic_equation_sllod_original_target_modified} can be
calculated as
\begin{equation}
 \label{dynamic_equation_sllod_original_target_modified_parallel_part}
 \begin{split}
   \frac{d\bm{\kappa}_{\parallel}(t)}{dt} \cdot \bm{r}(t) 
  & = \left[  \frac{d }{dt} 
  \left[ \bm{r}(t) \frac{\bm{r}(t) \cdot \bm{\kappa}(t) }{r^{2}(t)} \right]
  \right]\cdot \bm{r}(t) \\
  & =  \frac{d\bm{r}(t)}{dt} \frac{\bm{r}(t) \cdot \bm{\kappa}(t) \cdot \bm{r}(t)}{r^{2}(t)} 
   + (\text{terms parallel to $\bm{r}(t)$}) .
 \end{split}
\end{equation}
Here we utilize eq~\eqref{dynamic_equation_r_simple_sllod} and the identity $\bm{r}(t) \cdot \bm{\kappa}(t) \cdot \bm{r}(t) = \tr [\bm{r}(t) \bm{r}(t) \cdot \bm{\kappa}(t)]$.
Eq~\eqref{dynamic_equation_sllod_original_target_modified_parallel_part} can be rewritten as
\begin{equation}
 \label{dynamic_equation_sllod_original_target_modified_parallel_part_modified}
 \begin{split}
   \frac{d\bm{\kappa}_{\parallel}(t)}{dt} \cdot \bm{r}(t) 
  & =  \tr \bm{\kappa}_{\parallel}(t) 
  \left[ \frac{1}{m} \bm{p}_{\perp}(t) + \bm{\kappa}_{\perp}(t) \cdot \bm{r}(t) \right]
   + (\text{terms parallel to $\bm{r}(t)$}) \\
  & =  \frac{\tr \bm{\kappa}_{\parallel}(t) }{m} \bm{p}_{\perp}(t)
  + O(\| \bm{\kappa}(t) \|)
   + (\text{terms parallel to $\bm{r}(t)$}) .
 \end{split}
\end{equation}
The terms which are parallel to $\bm{r}$ can be absorbed into the Lagrange multiplier.
Thus we have eq~\eqref{dynamic_equation_sllod_original_target_modified} in the main text.

We show the calculations for eq~\eqref{shear_relaxation_modulus_dumbbells_overdamped}.
The correlation function for the bond vector can be analytically evaluated with
several different methods. Here we calculate $\langle r_{x}(t) r_{y}(t) r_{x}(0) r_{y}(0) \rangle_{\text{eq}}$
by using the Fokker-Planck equation \eqref{fokker_planck_equation_overdamped}.
We consider the situation where the initial bond vector is given as $\bm{r}(0) = \bm{r}_{0}$
and the external flow field is absent $\bm{\kappa}(t) = 0$.
Then eq~\eqref{fokker_planck_equation_overdamped} can be rewritten as
\begin{equation}
 \label{fokker_planck_equation_overdamped_explicit}
 \frac{\partial \psi(\bm{r},t)}{\partial t}
  = \frac{k_{B} T}{\zeta} \frac{\partial}{\partial \bm{r}} \cdot
  \left[ \left(\bm{1} - \frac{\bm{r}{\bm{r}}}{r^{2}}\right)
  \cdot \frac{\partial \psi(\bm{r},t)}{\partial \bm{r}} \right],
\end{equation}
and the initial condition is given as
\begin{equation}
 \psi(\bm{r},0) = \delta(\bm{r} - \bm{r}_{0}).
\end{equation}
The initial bond vector should obey the equilibrium distribution
\eqref{bond_vector_equilibrium_distribution}.
Then the correlation function can be rewritten as follows:
\begin{equation}
 \label{correlation_function_bond_vector}
  \begin{split}
   & \langle r_{x}(t) r_{y}(t) r_{x}(0) r_{y}(0) \rangle_{\text{eq}} \\
   & = \int d\bm{r}_{0}
    \left[ \int d\bm{r} \, r_{x} r_{y} \psi(\bm{r},t) \right] r_{0x} r_{0y} \psi_{\text{eq}}(\bm{r}_{0}) .
  \end{split}
\end{equation}
We calculate the integral over $\bm{r}$ in eq~\eqref{correlation_function_bond_vector},
$C_{xy}(\bm{r}_{0},t) \equiv \int d\bm{r} \, r_{x} r_{y} \psi(\bm{r},t)$.
From eq~\eqref{fokker_planck_equation_overdamped_explicit}, we have
\begin{equation}
 \begin{split}
  \frac{\partial C_{xy}(\bm{r}_{0},t)}{\partial t}
  & = \int d\bm{r} \, r_{x} r_{y} \frac{\partial \psi(\bm{r},t)}{\partial t} \\
  & = \frac{k_{B} T}{\zeta} \int d\bm{r} \, r_{x} r_{y} \frac{\partial}{\partial \bm{r}} \cdot
  \left[ \left(\bm{1} - \frac{\bm{r}{\bm{r}}}{r^{2}}\right)
  \cdot \frac{\partial \psi(\bm{r},t)}{\partial \bm{r}} \right] \\
  & = - \frac{k_{B} T}{\zeta} \int d\bm{r} \, \psi(\bm{r},t) \frac{\partial}{\partial \bm{r}} \cdot
  \left[ \frac{\bm{r}\bm{r}}{r^{2}}
  \cdot \frac{\partial (r_{x} r_{y})}{\partial \bm{r}} \right] \\
  & = - \frac{6 k_{B} T}{\zeta} \int d\bm{r} \,  \frac{r_{x}r_{y}}{r^{2}}\psi(\bm{r},t).
 \end{split}
\end{equation}
$r^{2}$ is constant during the time-evolution by the Fokker-Planck equation ($r^{2} = |\bm{r}_{0}|^{2} = b^{2}$), and thus we have
\begin{equation}
  \frac{\partial C_{xy}(\bm{r}_{0},t)}{\partial t}
   = - \frac{6 k_{B} T}{\zeta b^{2}} C_{xy}(\bm{r}_{0},t).
\end{equation}
Also, from the initial condition, we have $C_{xy}(\bm{r}_{0},t) = r_{0x} r_{0y}$.
Therefore we have the explicit expression for $C_{xy}(\bm{r}_{0},t)$ as
\begin{equation}
 \label{correlation_function_cxy_bond_vector}
 C_{xy}(\bm{r}_{0},t) = r_{0x} r_{0y} \exp\left( - \frac{6 k_{B} T}{\zeta b^{2}} \right).
\end{equation}
From eqs~\eqref{correlation_function_bond_vector} and \eqref{correlation_function_cxy_bond_vector},
finally we have
\begin{equation}
 \label{correlation_function_bond_vector_modified}
  \begin{split}
   & \langle r_{x}(t) r_{y}(t) r_{x}(0) r_{y}(0) \rangle_{\text{eq}} \\
   & = \exp\left( - \frac{6 k_{B} T}{\zeta b^{2}} \right) 
   \frac{1}{4 \pi b^{2}} \int d\bm{r}_{0} \,
r_{0x}^{2} r_{0y}^{2} \delta(|\bm{r}_{0}| - b) \\
   & =   \frac{b^{4}}{4 \pi}  \exp\left( - \frac{6 k_{B} T}{\zeta b^{2}} \right) 
 \int_{0}^{2 \pi} d\theta \int_{0}^{\pi} d\phi \, 
   \cos^{2} \theta \sin^{2} \theta \sin^{5} \phi \\
   & = \frac{b^{4}}{15} \exp\left( - \frac{6 k_{B} T}{\zeta b^{2}} \right) .
  \end{split}
\end{equation}
In eq~\eqref{correlation_function_bond_vector_modified} we have utilized the
variable transform from $\bm{r}_{0}$ to $\theta$ and $\phi$ defined via
$\bm{r}_{0} = [ b \cos \theta \sin \phi, \, b \sin \theta \sin \phi, \, b \cos \phi ]$.
By substituting eq~\eqref{correlation_function_bond_vector_modified} into eq~\eqref{shear_relaxation_modulus_dumbbells_overdamped},
we have eq~\eqref{shear_relaxation_modulus_dumbbells_overdamped_final} in the main text.


\end{document}